\documentclass{article}
\usepackage{graphicx}
\usepackage[utf8]{inputenc}
\usepackage{mathtools}

\let\subparagraph\relax
\usepackage{subfiles}
\usepackage{caption}
\usepackage{color}
\usepackage{amsmath}
\usepackage{pgfplots}
\usepackage{xspace}
\usepackage{authblk}
\usepackage{url}
\usepackage{listings}
\usepackage{xcolor}
\usepackage{graphics}

\usepackage[scaled]{beramono}
\usepackage[T1]{fontenc}

\title{DHT-based Edge and Fog Computing Systems: Infrastructures and Applications \thanks{© 2022 IEEE. Personal use of this material is permitted. Permission from IEEE must be obtained for all other uses, including reprinting/republishing this material for advertising or promotional purposes, collecting new collected works for resale or redistribution to servers or lists, or reuse of any copyrighted component of this work in other works.}
\thanks{978-1-6654-0926-1/22/18/\$31.00~\copyright2022 IEEE}
\thanks{Y. Hassanzadeh-Nazarabadi, S. Taheri-Boshrooyeh and Ö. Özkasap, "DHT-based Edge and Fog Computing Systems: Infrastructures and Applications," IEEE INFOCOM 2022 - IEEE Conference on Computer Communications Workshops (INFOCOM WKSHPS), 2022, pp. 1-6, doi: 10.1109/INFOCOMWKSHPS54753.2022.9798218.}}

\author{
Yahya Hassanzadeh-Nazarabadi, Sanaz Taheri-Boshrooyeh, 
and Öznur Özkasap\\
Department of Computer Engineering \\ 
Koç University, İstanbul, Turkey\\
\{yhassanzadeh13, staheri14, oozkasap\}@ku.edu.tr}

\begin{document}
\maketitle
\begin{abstract}
Intending to support new emerging applications with latency requirements below what can be offered by the cloud data centers, the edge and fog computing paradigms have reared. In such systems, the real-time instant data is processed closer to the edge of the network, instead of the remote data centers. With the advances in edge and fog computing systems, novel and efficient solutions based on Distributed Hash Tables (DHTs) emerge and play critical roles in system design. Several DHT-based solutions have been proposed to either augment the scalability and efficiency of edge and fog computing infrastructures or to enable application-specific functionalities such as task and resource management. This paper presents the first comprehensive study on the state-of-the-art DHT-based architectures in edge and fog computing systems from the lenses of infrastructure and application. Moreover, the paper details the open problems and discusses future research guidelines for the DHT-based edge and fog computing systems.
\end{abstract}

\section{Introduction}
\label{dht_survey_section_introduction}
Edge \cite{shi2016edge} and fog \cite{bonomi2014fog} computing are distributed system paradigms for supporting the latency-constrained applications that operate on instant real-time data, e.g., face recognition in smart phones or navigating the autonomous vehicles \cite{yi2015fog}. In such systems, the processing, computation and storage resources are moved away from the remote cloud data centers towards the vicinity of the edge of the networks, i.e., closer to the sources of the instant real-time data. The edge of a network is defined as the point where a local network interfaces the worldwide Internet, e.g., firewalls and routers. Computation and storage offloading to the edge and fog nodes instead of faraway cloud data centers reduce the amount of data on-move and the distance it must haul. Consequently, it provides a low-latency computing and storage platform for real-time applications. Compared to the traditional cloud computing paradigm, the edge and fog computing paradigms provide a low-latency and more efficient distributed data processing and storage infrastructure closer to the end-user. 
Hence edge and fog computing systems are more adaptable to the new emerging applications with latency requirements below what can be offered by the cloud computing systems, e.g., the $5^{th}$ Generation (5G)-based mobile network applications \cite{varshney2017demystifying}.

The edge and fog computing systems are composed of processes that are in constant need of communicating with each other and accessing each others' resources, e.g., data objects. The Distributed Hash Tables (DHTs) \cite{stoica2001chord} are considered as one of the prime solutions that address this need. DHT is distributed key-value store architecture of $n$ processes, where each process (i.e., node) maintains communication channels to $O(\log{n})$ other processes, which shapes a connected distributed overlay graph of processes. By utilizing this overlay graph, processes can maintain addressable entities (e.g., files) on DHT overlay, as well as query for each other and each others' entities, which resembles distributed \textit{put} and \textit{get} queries. Typically in DHTs, such distributed put and get queries are done within a message complexity of $O(\log{n})$. Due to their scalability, fault tolerance, fast searching, correctness under concurrency, and load balancing, DHTs are widely used in various edge and fog computing solutions to provide a scalable object storage platform \cite{sharma2020evaluation, simic2018edge, song2020smart, riabi2017proposal, sonbol2020edgekv, confais2017object}, as well as a resource discovery overlay \cite{tanganelli2018edge, tanganelli2017fog, santos2018towards, nakayama2017peer}. 

\textbf{Original Contributions:} To the best of our knowledge and compared to the relevant studies (e.g., \cite{yousefpour2019all, varshney2017demystifying, karagiannis2019compute, puliafito2019fog}), \textit{this is the \textbf{first study} that proposes a taxonomy of the state-of-the-art DHT-based edge and fog computing solutions focusing on their system architecture, routing, and technological aspects as well as identifies advanced and significant capabilities empowered by DHTs}. Our taxonomy goes beyond the commonly known DHT methods such as storage, routing, and lookup, and extracts sophisticated and diverse DHT-based solutions including but not limited to two-tier overlays and storage systems, content distribution, intrusion detection, and resource discovery and management. Furthermore, we identify open problems and discuss future research guidelines for DHT-based edge and fog computing systems.

\textbf{Selection Criteria and Taxonomy:} We exclude the dated publications (i.e., $2015$ and prior) and look into the recent publications (i.e., $2015-2021$) that yield a high correlation with the keywords set of \textit{DHT, Edge, Fog, and Cloudlet}. We investigate the resulting publications and shortlist only those in which DHT utilization plays a critical role in the performance, correctness, or reliability of the proposed solution. Based on their relevance, we further divide the shortlisted solutions into either of the following categories: \textit{Infrastructure} or \textit{Application}. The infrastructure category represents solutions that utilize DHT to augment the scalability and efficiency of the edge and fog computing platforms regarding the routing \cite{d2018sa} and data object storage \cite{sonbol2020edgekv}. The application category, on the other hand, represents the solutions which utilize DHTs to enable an application-specific functionality on top of an edge or fog computing platform, e.g., access control management \cite{riabi2017proposal}, content distribution  \cite{nakayama2017peer}, service \cite{santos2018towards} and resource \cite{tanganelli2017fog} discovery, task delegation  \cite{gedeon2017router} and management \cite{simic2018edge}, resource management  \cite{song2020smart}, and intrusion detection \cite{sharma2020evaluation}.

\section{System Model}
\label{edge_survey_section_system}
\begin{figure}
\begin{center}
    \includegraphics[scale=0.20]{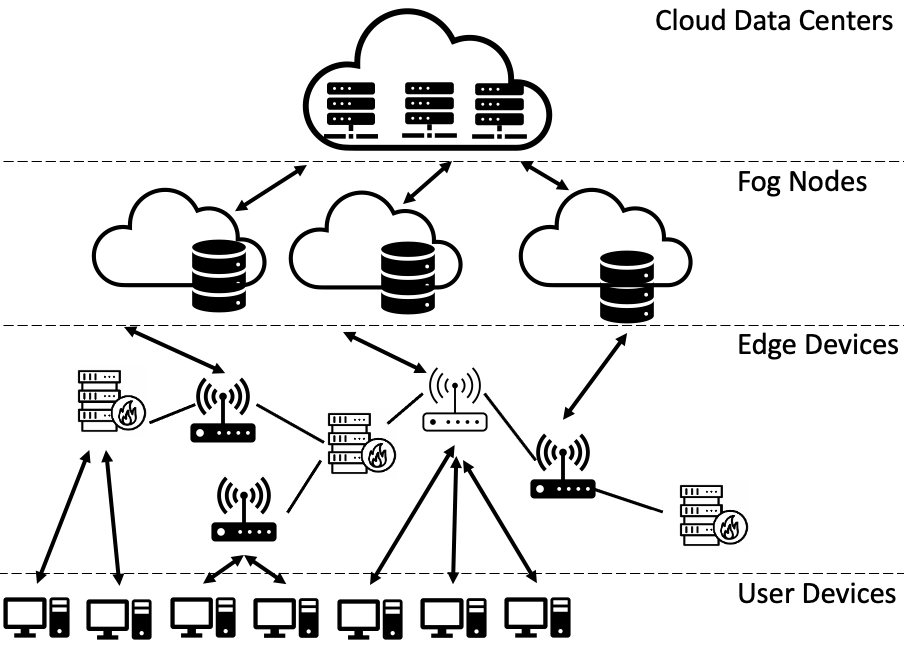}
\caption{An overview of system model in the edge and fog computing paradigm.} \label{fig_edge_model}
\end{center}
\end{figure}

\textbf{Edge and Fog Computing:} As illustrated by Figure \ref{fig_edge_model}, the edge and fog computing paradigms are organized into a system of four layers, namely: the user devices, edge devices, fog nodes, and cloud data centers. The edge devices are resource-constrained devices (e.g., routers) that are directly connected to the user devices in their local network, and are responsible for providing computation and storage power for the latency-constrained operations \cite{dolui2017comparison}. Due to their resource limitations running multiple real-time applications on edge devices can easily lead them to resource contention \cite{dastjerdi2016fog}. This is where the fog computing paradigm comes into play. Fog computing is defined as the cluster of heterogeneous resourceful devices (e.g., high-end servers) that are interconnected with high bandwidth and placed one hop away from the edge of the network \cite{varshney2017demystifying}. Fog nodes are placed between the edge and cloud data centers layers and support the edge layer by providing a runtime platform with predictable low latency. Fog nodes also add more decentralization by moving the real-time processing and storage power down from the cloud data centers to the replicated nodes closer to the edge devices \cite{bonomi2012fog}. In the edge and fog computing system model, the cloud data centers are utilized primarily to support the fog nodes by processing and archiving aggregated non-latency-constrained data provided by fog nodes at a larger scale.

\textbf{Distributed Hash Tables (DHTs):} A DHT \cite{stoica2001chord} is a distributed overlay graph of connected processes, i.e., \textit{DHT nodes}. The DHT nodes utilize the overlay graph to efficiently search for each other as well as each others' resources, e.g., data objects. In such overlay graphs, each process is represented by a DHT node identified by a unique \textit{identifier} on the overlay DHT graph. The identifiers of the DHT nodes are determined by taking the collision-resistant \cite{katz2014introduction} hash value of their unique overlay network (IP) address or their cryptographic public key. The resources of DHT nodes are also represented by unique identifiers, which are typically determined using the collision-resistant hash value of their content. 
In a DHT overlay with $n$ nodes, a DHT node commonly maintains connections to $O(\log{n})$ other nodes in a table, which is called its \textit{lookup table (i.e., routing table)} \cite{hassanzadeh2020interlaced}. 
The lookup table neighbor relationship forms a connected overlay graph of processes, which enables them to store and maintain their resources in the DHT key-value store, as well as to efficiently search for each other and each others' resources across the DHT overlay. Both operations are done in a fully decentralized and distributed manner within a message complexity of $O(\log{n})$. Storing a resource (e.g., a data object) in a DHT is a distributed \textit{put} operation that lays that resource in the DHT node that has the closest identifier to it. The identifier closeness is domain-specific for each DHT and is defined in the identifier space of that DHT. A lookup operation corresponds to a distributed \textit{get} operation, which is initiated by a \textit{lookup initiator} node for a \textit{target identifier}. The target identifier can be the identifier of a DHT node (i.e., process) or a resource. 
As the result of a lookup operation, the lookup message is routed collectively by the nodes on the lookup path from the lookup initiator to the node with the closest identifier to the lookup target, which terminates the lookup and announces itself as the lookup result back to the initiator.

\section{DHT-Based Infrastructure Solutions}
\label{edge_survey_section_infrastructure}
\subsection{Routing Overlays}
\label{edge_survey_subsection_routing_overlays}
\begin{figure}
\begin{center}
    \includegraphics[scale=0.20]{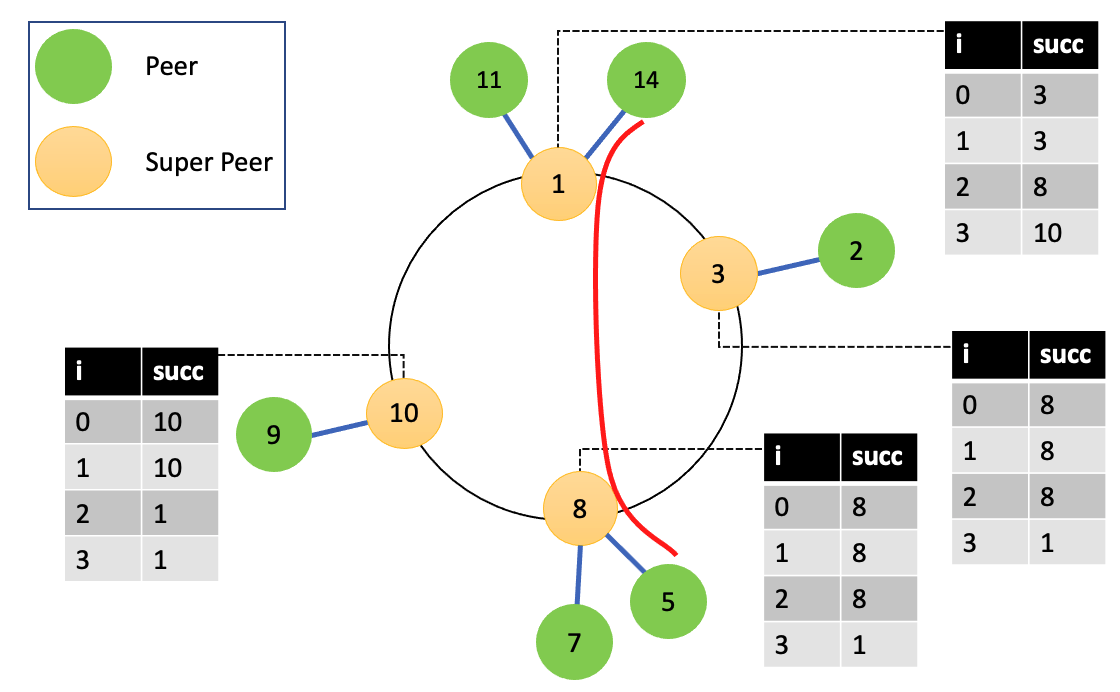}
\caption{An example of two-tier DHT overlay structures \cite{d2018sa} for heterogeneous edge computing systems. The lookup tables of super-peers on are represented by grey tables connected to them. An example of a lookup path from regular peer $14$ to regular peer $5$ is illustrated by the red line. The lookup is routed through their corresponding super peers.} \label{fig_edge_sachord}
\end{center}
\end{figure}

\textbf{Two-tier Overlays \cite{d2018sa}:} Edge computing relies on decentralized overlay networks as the communication backbone among the heterogeneous resource-constrained edge devices. Involving all these edge devices in the overlay routing functionality may impose runtime uncertainty and degrade the reliability of the system. To tackle this issue, the two-tier DHTs of super-peers and regular peers are utilized, which take into account the heterogeneity of the resources of the edge devices \cite{d2018sa}. An example of such two-tier topology overlays with $4$ super-peers and $6$ regular peers is shown in Figure \ref{fig_edge_sachord}. Edge devices in such two-tier DHT-based routing overlays are assigned either as the \textit{regular peers} or \textit{super-peers}. The decision on being a super-peer or a regular peer depends on the edge device's available resources compared to the average available resources of the entire system, which is aggregated in a distributed way. For this sake, there is an objective function that evaluates peer's resources to a numerical value. Each peer computes the objective value locally based on its available resources, compares it against the aggregated average objective value of the network, and determines on being a regular peer or a super-peer accordingly. The objective function is designed such that the edge devices determined as \textit{super-peers} are resourceful strong peers, while the \textit{regular peers} are the resource-restricted ones. The core routing functionality in such two-tier DHTs is established over the super-peers, which route queries between each other as well as the regular peers, while the regular peers are only the consumers of this infrastructure and exchange their queries through super-peers. For this sake, the super-peers establish a DHT-based routing overlay among themselves, where each super-peer maintains a routing table of a subset of other super-peers (e.g., based on the Chord protocol \cite{stoica2001chord} in SA-Chord \cite{d2018sa}). Each regular peer is assigned to the super-peer that immediately succeeds it on the identifier space, i.e., the super-peer with the smallest identifier that is greater than the regular peer's identifier. 

\subsection{Object Storage }
\textbf{Two-tier Storage \cite{sonbol2020edgekv}:} In addition to their reliable routing functionality, the two-tier DHTs are also utilized to establish strongly-consistent fault-tolerant object storage among edge devices. As mentioned in Section \ref{edge_survey_subsection_routing_overlays}, in such two-tier DHTs the edge devices are categorized as regular peers and super peers. Each group of regular peers is assigned to a super-peer that represents the group on the DHT overlay of super-peers. The messages to and from a group of regular peers are routed using their super-peer. In the two-tier DHT-based storage systems, the regular peers of each group form a complete overlay graph among themselves and are responsible for storing the same set of data objects, i.e., key-value pairs. A user submits her data object read/write query to the DHT overlay of super-peers which is routed to the group of regular peers responsible for the data object's key. The regular peers of a group form a replicated state machine \cite{bessani2014state} of key-value pairs assigned to that group. A write request gets through a consensus and is written on all regular peers of the group. This results in a consistent storage state across all the servers of each group. Therefore, a read request for a key can be handled by any of the regular peers in that group. The replicated state machine-style for reads and writes is also fault-tolerant, i.e., all regular peers keep the same state of data objects, hence failing some of them does not affect the data availability \cite{hassanzadeh2016awake} of the system.

\begin{figure}
\begin{center}
    \includegraphics[scale=0.3]{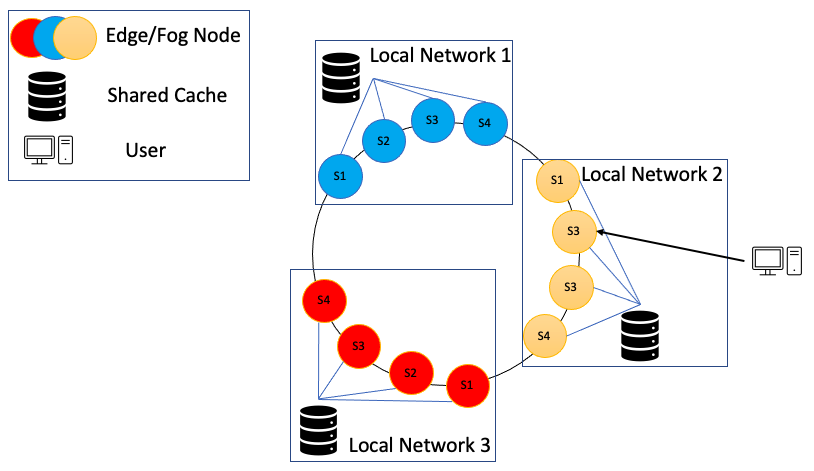}
\caption{Sample architecture of the cached-based storage for DHT overlays of edge and fog nodes \cite{confais2017object}. The system is comprised of independent local networks. The edge and fog nodes on each local network are nodes on a global DHT and share the same cache instance.} \label{fig_edge_nas}
\end{center}
\end{figure}

\textbf{Cache-based Storage \cite{confais2017object}:} The access latency of retrieving data objects in the DHT-based edge and fog storage systems can be further improved by enabling DHT nodes of the same local network with a shared cache. The query results made by a DHT node are maintained on the shared cache. Upon receipt of a data query from a user, the DHT node looks up the shared cache first before trying a lookup on the DHT overlay. It is trivial that resolving a query through the shared cache on the local network is several orders of magnitude faster than resolving through DHT lookups that go through several remote nodes. Figure \ref{fig_edge_nas} illustrates an example of such system architectures. Such shared caching solutions can be optimized even further by treating data objects immutable, hence the caching does not require any invalidation.

\section{DHT Applications in Edge/Fog Computing}
\label{edge_survey_section_application}
\begin{table}
\centering
\resizebox{\columnwidth}{!}{%
    {
        \begin{tabular}{ |l|l|l|l|l|l|l|l|  }
        \hline
        Solution &
        Type &
        DHT &
        Nodes & 
        Identifiers & 
        DHT Utilization &
        Domain \\
        \hline
        Two-tier Overlay \cite{d2018sa} & 
        Inf &
        Chord & 
        Super Peers & 
        - & 
        Routing Overlay & 
        Decentralized \\
        Two-tier Storage \cite{sonbol2020edgekv} &
        Inf & 
        Chord &
        Super Peers &
        Content &
        Object Storage &
        Decentralized \\
        Cache-based Storage \cite{confais2017object} &
        Inf & 
        Kademlia &
        Data Objects &
        Content &
        Object Storage &
        Decentralized \\
        Content Distribution  \cite{nakayama2017peer, gedeon2017router} &
        App &
        Chord &
        Edge servers &
        IP address &
        Routing Overlay &
        Distributed \\
        Access Control Management \cite{riabi2017proposal} &
        App &
        Chord &
        Access Control Lists &
        - &
        Object Storage &
        Decentralized \\
        Service Discovery \cite{santos2018towards} &
        App &
        General &
        Services &
        Content &
        Routing Overlay &
        Decentralized \\
        Resource Discovery (Mobile Environments) \cite{tanganelli2017fog} &
        App &
        General &
        Resources &
        Attribute Values &
        Routing Overlay &
        Decentralized \\
        Resource Discovery (Unreliable Environments) \cite{tanganelli2018edge} &
        App &
        General &
        Devices &
        URI &
        Routing Overlay &
        Distributed \\
        Producer-Consumer Buffer  \cite{song2020smart} &
        App &
        General &
        Data Objects &
        Content &
        Object Storage &
        Decentralized \\
        Collaborative Computation  \cite{simic2018edge} &
        App &
        General &
        Computation Results &
        Content &
        Object Storage &
        Distributed \\
        Intrusion Detection \cite{sharma2020evaluation} &
        App &
        Kademlia &
        SYN Packets' History &
        IP address &
        Object Storage &
        Decentralized \\
        \hline
        \end{tabular}
    }
    }
    \caption{A summary of DHT-based solutions in Edge/Fog computing. \textit{Inf} and \textit{App} are abbreviations for \textit{Infrastructure}, and \textit{Application}, respectively.}
    \label{dht_survery:table_edge_comparison}
\end{table}

\subsection{Routing Overlays}
\textbf{Efficient Content Distribution \cite{nakayama2017peer, gedeon2017router}:}
In a DHT overlay of $n$ nodes, each node maintains persistent connections to $O(\log{n})$ nodes, i.e., its neighbors. The neighboring relationship in DHTs yields a connected overlay graph with both the degree of nodes as well as the graph diameter equivalent to $O(\log{n})$. In the DHT-based edge computing overlays, the overlay graph is also utilized to efficiently broadcast content from a content owner to all other participating nodes. Such content distribution through broadcasting takes $O(\log{n})$ dissemination time and $O(n \times \log{n})$  worst-case message complexity. An example of such solutions is detailed in \cite{nakayama2017peer} where a DHT-based overlay of the edge servers is utilized to provide video streaming through content distribution. Each user is both a content receiver as well as a content distributor and is connected to an edge server on the DHT overlay. The content distribution is done by each edge server broadcasting the video streams of its users to its DHT neighbors who similarly broadcast to their neighbors. Thereby, the users' video streams eventually reach all DHT nodes, hence made available for all users in asymptotic time complexity of $O(\log{n})$. 

\textbf{Scalable Service and Resource Discovery \cite{santos2018towards, tanganelli2018edge}:} Fog computing systems are distributed in their nature and inherently dispersed geographically. Nevertheless, the runtime services and resources they offer (e.g., computing services and storage buckets) can benefit the users globally. By registering their resources and services on a global DHT overlay, the fog nodes make them available for users to discover through DHT lookups in a decentralized manner. Such decentralized service and resource discovery is experimentally proven more efficient and scalable than the centralized cloud-based solutions \cite{santos2018towards}. Fog nodes form a DHT overlay and register their services as key-value pairs, where the key is the service identifier, and the value is the service template that contains all information to mount, register, and instantiate the service.
In such system models, the user devices are bootstrapped with the network address of a subset of the DHT fog nodes, which enables them to query the DHT through those fog nodes.

\textbf{Range Queries \cite{tanganelli2017fog}:} The service and resource discovery offered by DHT overlays can be further evolved to support range queries by utilizing two-tier DHTs (see Section \ref{edge_survey_subsection_routing_overlays}). In such structures, each tier-one DHT node (i.e., super-peer) is a fog node, which represents a resource type (e.g., a computing engine), and is attached to a tier-two DHT of resource nodes (i.e., regular nodes). The identifier of a fog node in the tier-one DHT is the resource name it represents. The identifier of a resource node in the tier-two DHT is the vector of its attributes' values, e.g., available CPU cores, storage, memory, etc. Looking up resources with a specific range of attributes is done in two steps. 
In the first step, the query is dispatched in the tier-one DHT to look up the fog server responsible for that resource type. Then the tier-two DHT attached to the responsible fog server is queried for the desired range of the attributes it represents.
As the result, the resource instances with the desired range of attribute values are returned.

\subsection{Object Storage}
\textbf{Shared Object Storage \cite{riabi2017proposal, lemon2002resisting}:}
As a decentralized key-value store, the DHT overlays are utilized to establish scalable storage solutions for sharing data objects among participating edge or fog nodes. Objects in such storage platforms are modeled as key-value pairs, where \textit{key} is the unique identifier of the object and is utilized for DHT-based storage and retrieval of the object. The \textit{value} represents the actual object to be stored. An example of such decentralized storage of shared objects is maintaining the Access Control List (ACL) of transportation resources (e.g., navigation services) over DHT of fog nodes  \cite{riabi2017proposal}. An ACL embodies the list of resources and their access permissions. Maintaining ACLs over DHT of fog nodes establishes a fine-grained and scalable access control management for transportation resources. In such systems, users are smart vehicles that are commuting between fog nodes across the city. The transportation resources are managed by the fog nodes on the DHT overlay. A user aiming to exploit a transportation resource submits an access request to the closest fog node. The access request is translated into a lookup over the DHT overlay of fog nodes to fetch the ACL of the requested resource, and accordingly, the request of the user is evaluated. A granted request is done by generating an access token of the requested resource for the user. The other example for DHT-based storage of shared objects is to build up an intrusion detection system among edge server nodes as a defense mechanism against SYN flooding denial of service attacks \cite{lemon2002resisting}. The SYN floods are a specific type of denial of service attack that aim at exhausting the resources of a victim server so that it can not respond to legitimate queries in a timely fashion. The attack exploits a vulnerability in the handshake phase of TCP connections. The handshake phase starts with one party sending a SYN packet to a server, which allocates some connection-related resources on the server-side. In the SYN flood attacks, the attacker starts sending a flood of SYN packets to the victim server, where each packet corresponds to a distinct TCP connection request (e.g., originating from distinct ports of the attacker). Since each packet causes the victim to allocate some resources for the prospective TCP connection, the flood of such packets exhausts the communication resources of the victim. To defend themselves against such SYN flooding attacks at the edge of the network, the edge servers establish a DHT overlay and maintain the SYN packet history they receive over the DHT \cite{lemon2002resisting}. The key of a data object on this DHT is the origin IP address of the SYN packet, and the value is the SYN request history corresponding to that address. Upon receiving each incoming SYN packet, an edge server updates the value corresponding to the packet's origin IP address over the DHT by appending the information of the new packet. Any incoming SYN packet belonging to a new TCP handshake request is then checked by the edge servers to see whether the SYN packets history originated from the requesting IP address within a certain time interval exceeds a threshold. If it holds, then the source IP address is marked as malicious, and the edge node drops its allocated resources for that. Sharing the records of SYN packets over the DHT in this solution allows the edge servers to collaboratively identify and deter the SYN flood attacks.

\textbf{Producer-Consumer Buffer \cite{song2020smart, simic2018edge}:} Edge computing systems are comprised of resources that provide storage and computing services to the end-users. The computation providers typically operate on the data objects that are maintained by the storage provider edge devices. From this perspective of operating on data, the computation provider edge devices are seen as the \textit{consumer} of the data objects that are \textit{produced} by the users and \textit{maintained} over the storage provider edge devices. Thereby, the storage provider nodes are presumed as a buffer zone between the producer and consumer entities in this system model. In such scenarios, the DHT overlay of storage provider nodes is utilized as a decentralized, persistent, efficient, and scalable buffer between the producer and consumer entities. This makes the data objects efficiently accessible by the computation providers, to fetch and operate on them. For example in \cite{song2020smart}, the end-users (i.e., producers) submit a computation task to a smart contract on the Ethereum blockchain \cite{wood2014ethereum}, which determines the storage and computation provider edge devices for that task. Once the assignment of the computation and storage providers are done, the user breaks the computation task into chunks and stores them over the DHT of the storage providers, which makes them accessible by the computation providers. Computation providers then retrieve the chunks from the storage DHT and execute the task. A similar solution is also proposed in \cite{simic2018edge}, where DHT of storage providers is utilized to maintain the interim computation results of users' sub-tasks. The system model of this solution is composed of clusters of storage and computation provider edge devices, where each cluster is managed by a cluster master. The submitted computation task of the user is broken into chunks by the cluster master and distributed among the computation providers. For a certain task, the sub-tasks should be executed in a specific order of precedence, i.e., the execution of each sub-task is done on top of the result of the preceding sub-tasks. Hence, the interim computation results are stored on the DHT of storage providers within that cluster and are made available for the subsequent sub-tasks.

Table \ref{dht_survery:table_edge_comparison} summarizes the DHT-based edge and fog computing solutions. In this table, the \textit{Nodes} column refers to the primary representation of the nodes once a DHT overlay of physical devices (e.g., servers) is shaped. For example, in the DHT-based task management application \cite{simic2018edge} once the DHT overlay is shaped among the physical devices, it is primarily utilized for storing computation results as the DHT nodes. 
Also, the \textit{Identifiers} column represents the (unique) attribute of the DHT nodes that is used as the input to a collision-resistant hash function \cite{katz2014introduction} for generating their unique identifiers. The \textit{DHT Utilization} column represents the primary purpose of DHT utilization in the solution. The \textit{Domain} column represents how the DHT nodes are controlled by the administrative domain(s). In a centralized solution, all nodes are managed by a single monolith administrative domain. In a distributed solution, nodes are managed by several agents all belonging to the same administrative domain. In a decentralized solution, the nodes are partitioned among several independent administrative domains, where each domain controls its subset of nodes.

\section{Open Problems and Research Guidelines}
\label{edge_survey_open_problems}

\textbf{Query Load Balancing:} 
The query load balancing is the problem of managing the load of request \textit{queries} received by DHT nodes (i.e., edge/fog servers) in a decentralized way. The edge and fog servers running as DHT nodes typically also act like gateways providing an interface for the external users to query the DHT overlay \cite{riabi2017proposal, santos2018towards}. Since each user is normally connected to the closest DHT node (i.e., server) in the underlying network, servers in the more congested areas undertake a higher query load compared to the sparse regions. This results in a non-uniform available bandwidth and quality-of-service distribution across the system, which degrades the performance of the system in crowded locations and increases the failure likelihood. Balancing the request load of edge and fog servers involved in a DHT overlay protocol is a potential research direction \cite{hassanzadeh2019decentralized}.

\textbf{Decentralized Stabilization:}
A fundamental challenge in DHT-based solutions is the cost associated with the lookup table stabilization in the presence of nodes' failure as well as their dynamic online and offline states. Many studies \cite{gedeon2017router, nakayama2017peer, tanganelli2018edge} opt-in centralized solutions where a central registry server configures DHT nodes (i.e., edge and fog servers) with their DHT lookup tables and keeps updating the tables following their dynamic behavior. Efficient and decentralized DHT stabilization algorithms \cite{hassanzadeh2020interlaced} should be further studied and integrated into the edge and fog computing ecosystems.

\textbf{Range Queries:} 
A range query on the DHT overlays is typically done through individual queries for each value in the range \cite{tanganelli2017fog}. This imposes $O(k \log{n})$ message complexity to resolve a range query spanning $k$ values in a system with $n$ DHT nodes. If $k$ outnumbers $n$ then the range query causes linear message complexity. Developing efficient DHT-based range query techniques \cite{zheng2006distributed} for edge and fog computing ecosystems can significantly benefit system performance and should be studied further.

\textbf{Malicious Behavior:}
The existing edge and fog computing systems rely on the trustworthiness of participants to correctly follow the protocols and deliver their assigned functionality. Such solutions disregard the potential malicious behavior of participants and the subsequent performance, service reliability, and security harms. For example, in delegated computations like \cite{song2020smart}, a malicious resource provider may provide junk computational results in favor of saving its resources, and the users which are resource-constrained entities are not capable of verifying the output through the re-computation. Similar misbehavior may allow malicious super-peers to get a free ride of the system by pretending to be a regular peer in the two-tier overlays \cite{d2018sa}. Malicious nodes can also attack the availability and integrity of DHT-based routing by deviating from the routing protocol, e.g., falsifying the result or dropping messages they should route. The verifiable protocol execution \cite{goldwasser2015delegating} and secure routing techniques \cite{boshrooyeh2017guard} should be studied further for the integration into the current fog and edge computing environments.

\textbf{Super vs Regular Peers Ratio:} 
As studied in this paper, the two-tier routing and storage overlays are extensively utilized in the edge and fog computing systems \cite{d2018sa, tanganelli2017fog}. The current solutions lack finding the equilibrium point expressing the threshold ratio of super-peers to the regular peers in the system. Addressing this equilibrium point is crucial for the performance of the system in satisfying its objectives. Super-peers outnumbering the regular peers results in under-utilization of available resources, while the reverse results in overloading the super-peers and leading them to failure. Finding the equilibrium point between the super and regular peers in the two-tier DHT architectures considering the dynamic behavior of nodes is a potential research direction and should be studied further.

\textbf{Metrics Aggregation:} The critical performance-related decisions of the surveyed solutions are typically taken based on an average value of the aggregated data, e.g., the decision on being a regular vs super peer in the two-tier DHTs \cite{d2018sa, tanganelli2017fog}. However, the aggregated average value is not necessarily the best option to decide upon especially in situations where the standard deviation is notably low or high, which yields in noisy decisions \cite{kleppmann2017designing}, e.g., the majority of nodes being selected as peers or super peers in the two-tier architectures. Investigating more indicative ways of aggregating the performance metrics (e.g., median) for edge and fog computing solutions is a potential research direction and should be studied further.

\section{Conclusion}
As a distributed key-value store architecture with asymptotic logarithmic operational complexity, Distributed Hash Tables (DHT) are widely used in edge and fog computing solutions to address challenges such as scalability, availability, reliability, and performance. In this paper, we presented the first taxonomy study on state-of-the-art DHT-based edge and fog computing solutions from the system architecture, routing, and technological perspectives. Our taxonomy concludes that in addition to their fundamental application as two-tier routing and storage architectures, DHTs are increasingly utilized for applications like efficient content distribution, scalable service, and resource discovery, range queries, and shared object storage platforms for a variety of resources ranging from simple plain data objects to intermediary computation results and access control lists. Furthermore, we identified open problems and discussed future research directions for the DHT-based edge and fog computing solutions considering their performance, decentralization, experimentation, and security.

\bibliographystyle{IEEEtran}
\bibliography{references}
\end{document}